\documentclass[aps,prl,notitlepage,twocolumn,10pt]{revtex4-1}
\usepackage{graphicx}
\usepackage{bm}
\usepackage{amsfonts,amsmath}
\usepackage[colorlinks=true]{hyperref}

\begin{document}

\title{Viscous dynamics of vortices in a ferromagnetic film}

\author{Derek Reitz}
\author{Anirban Ghosh}
\author{Oleg Tchernyshyov}
\affiliation{
Department of Physics and Astronomy, 
Johns Hopkins University,
Baltimore, Maryland 21218, USA
}

\begin{abstract}
We derive viscous forces for vortices in a thin-film ferromagnet. The viscous force acting on vortex $i$ is a linear superposition $\mathbf F_i = - \sum_{j} \hat{D}_{ij} \mathbf V_j$, where $\mathbf V_j$ is the velocity of vortex $j$. Thanks to the long-range nature of vortices, the mutual drag tensor $\hat{D}_{ij}$ is comparable in magnitude to the coefficient of self-drag $D_{ii}$. 
\end{abstract}

\maketitle

The dynamics of solitons in ferromagnets is a topic with a long history. Time evolution of magnetization, represented by the field $\mathbf m(\mathbf r,t)$ of unit length, is described by the Landau-Lifshitz equation 
\begin{equation}
\mathcal J \dot{\mathbf m} = - \mathbf m \times \frac{\delta U}{\delta \mathbf m} - \alpha |\mathcal J| \mathbf m \times \dot{\mathbf m},
\label{eq:LLG}
\end{equation}
where $U[\mathbf m(\mathbf r)]$ is a functional of potential energy, $\mathcal J$ is the density of angular momentum \cite{sign-convention}, and $\alpha \ll 1$ is Gilbert's damping constant \cite{Gilbert2004}. Even in the simplest models, where the energy includes only exchange interactions and local anisotropy, Eq.~(\ref{eq:LLG}) is a nonlinear partial differential equation that rarely admits exact dynamical solutions. Approximate solutions can be obtained for soft modes associated with global symmetries (such as translations) in the limit of weak external perturbations. \textcite{Thiele1973} described the dynamics of a rigidly moving magnetic soliton, $\mathbf m(\mathbf r - \mathbf R(t))$, whose velocity $\dot{\mathbf R}$ is determined from the equation
\begin{equation}
\mathbf G \times \dot{\mathbf R} - \partial U/\partial \mathbf R - D \dot{\mathbf R} = 0,
\label{eq:Thiele}
\end{equation}
expressing the balance of gyroscopic, potential, and viscous forces, respectively. 

\begin{figure*}[bt]
\includegraphics[width=1.9\columnwidth]{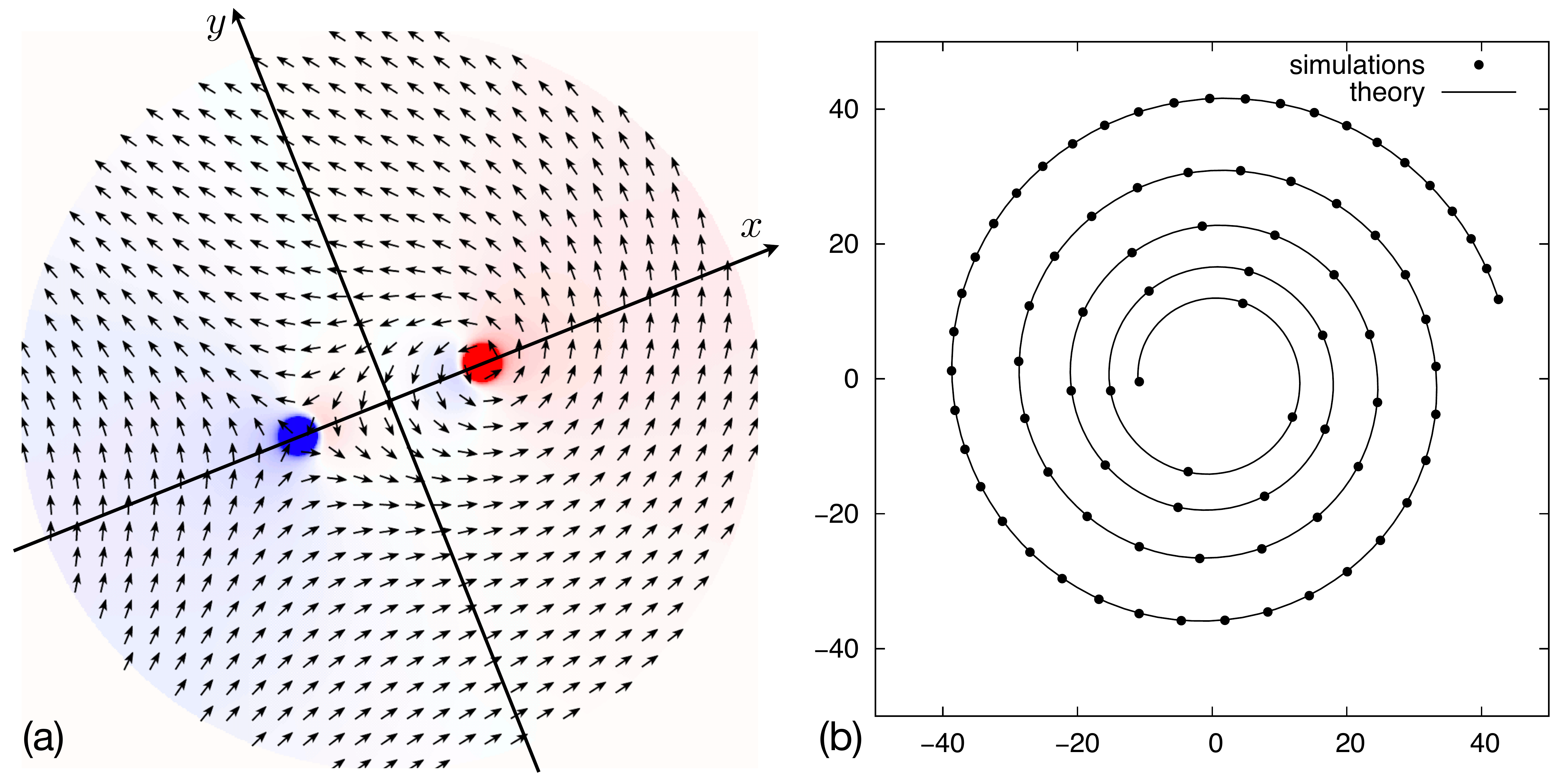}
\caption{(a) Vortex--antivortex pair with equal skyrmion numbers. Red and blue colors signify positive and negative out-of-plane magnetization $m_z$, respectively. (b) Trajectory of the vortex. Natural units of length.}
\label{fig:v-av}
\end{figure*}

Thiele's equation (\ref{eq:Thiele}) has been widely used to describe the dynamics of vortices in a thin film \cite{Takeno1982, Huber1982, Nikiforov1983, Mertens1997, Mertens2000, Gaididei2000, Guslienko2002, Kovalev2002, Buchanan2005, Sheka2005, Komineas2007, Ivanov2010}. With few exceptions \cite{Huber1982, Mertens2000, Sheka2005}, analytical treatments take into account the gyroscopic and potential forces but leave out the viscous force involved in energy dissipation. Both the gyroscopic and viscous forces are proportional to the soliton velocity $\dot{\mathbf R}$ and the neglect of the viscous force can be justified by its relative weakness: the viscosity tensor $D_{ab} = \alpha |\mathcal J| \int dV \, \partial_a \mathbf m \cdot \partial_b \mathbf m$ is of a higher order in $\alpha \ll 1$ than the gyrovector $G_a = \epsilon_{abc} \mathcal J \int dV \, \mathbf m \cdot (\partial_b \mathbf m \times \partial_c \mathbf m)$. Nonetheless, in certain situations the viscous force cannot be neglected. For example, the annihilation of a vortex and an antivortex is accompanied by gradual dissipation of energy as the two solitons approach each other. This motivates us to seek a proper understanding of viscous forces in vortex dynamics. 

To be specific, we set as our immediate goal to obtain a satisfactory analytical model for the motion of a vortex--antivortex pair with equal skyrmion numbers, Fig.~\ref{fig:v-av}(a). The two solitons attract each other through a potential force mediated by exchange interaction. To the zeroth order in $\alpha$, Thiele's equation (\ref{eq:Thiele}), applied to each soliton separately, predicts that they will orbit a common center at an orbital velocity proportional to the force of attraction. At the next order in $\alpha$, viscous forces opposing the orbital motion will induce slow radial motion of the solitons toward each other, Fig.~\ref{fig:v-av}(b). 

Our main findings are as follows. (1) Viscous forces acting on vortices come in two kinds. The first is self-drag, a force proportional to the vortex's own velocity \cite{Huber1982, Mertens2000}. We show that a vortex also experiences a drag force from other vortices proportional to \emph{their} velocities. The net force of viscous friction acting on vortex $i$ is $\mathbf F_i = - \sum_{j}\hat{D}_{ij} \mathbf V_j$, where $\mathbf V_j$ is the velocity of vortex $j$ and $\hat{D}_{ij}$ is the mutual drag tensor comparable in magnitude to the self-drag coefficient $D_{ii}$. (2) Both $D_{ii}$ and $\hat{D}_{ij}$ scale logarithmically with the system size.  (3) The direction of mutual drag depends on the product of vorticities. A vortex receding from an antivortex attempts to drag the antivortex with it; the force direction reverses for a vortex--vortex pair. (4) Image vortices, created by ``reflection'' in the sample edge, produce substantial corrections to viscous forces.

Our theory is built on the framework of collective coordinates \cite{Mertens2000, Tretiakov2008,Clarke2008}, in which the magnetization field $\mathbf m(\mathbf r,t)$ is parametrized by a few coordinates $\{q^\mu\}$, $\mu = 1, 2, \ldots$, representing soft modes of the system (e.g., vortex positions). The Landau-Lifshitz equation (\ref{eq:LLG}) translates into equations of motion for each coordinate $q^\mu$, 
\begin{equation}
G_{\mu \nu} \dot{q}^\nu - \frac{\partial U}{\partial q^\mu} - D_{\mu \nu} \dot{q}^\nu = 0.
\label{eq:CC}
\end{equation}
Thiele's equation (\ref{eq:Thiele}) is a particular case of Eq.~(\ref{eq:CC}), in which $\{q^\mu\}$ are global translations $\mathbf m(\mathbf r) \mapsto \mathbf m(\mathbf r - \mathbf R)$. The gyroscopic and dissipative tensors are \cite{Tretiakov2008,Clarke2008} 
\begin{subequations}
\begin{eqnarray}
G_{\mu \nu} &=& - \mathcal J \int dV \, 
	\mathbf m \cdot 
    \left( 
    	\frac{\partial \mathbf m}{\partial q^\mu}
        \times
        \frac{\partial \mathbf m}{\partial q^\nu}
    \right),
\label{eq:G-Da}
\\
D_{\mu \nu} &=& \alpha |\mathcal J| \int dV \, 
    	\frac{\partial \mathbf m}{\partial q^\mu}
        \cdot
        \frac{\partial \mathbf m}{\partial q^\nu}.
\label{eq:G-Db}
\end{eqnarray}
\label{eq:G-D}
\end{subequations}

We use a simple model of a thin-film ferromagnet with exchange interaction of strength $A$ and easy-plane anisotropy of strength $K$. We omit dipolar interactions \cite{dipolar}. In a film of thickness $h$, the energy is
\begin{equation}
U = h \int d^2r
	\left(  
    	A |\nabla \mathbf m|^2 
        + K m_z^2
    \right)/2.
\end{equation}

The polar angle of magnetization $\theta$ is a hard mode pinned at $\theta = \pi/2$. At low energies, the system is effectively an $XY$ ferromagnet \cite{Chaikin-Lubensky} parametrized by the azimuthal angle of magnetization $\phi(\mathbf r,t)$ with energy $U =  h \int d^2r \, A |\nabla \phi|^2/2$. A state with $N$ vortices has
\begin{equation}
\phi(\mathbf r) = \sum_{i=1}^N n_i \arctan{\frac{y-Y_i}{x-X_i}},
\label{eq:phi-vortices}
\end{equation}
where $\mathbf R_i = (X_i, Y_i)$ is the location of the $i$th vortex and $n_i \in \mathbb Z$ is its vorticity, usually $n_i = \pm 1$. The effective description breaks down inside vortex cores---circular regions with the size on the scale of the exchange length $\lambda =\sqrt{A/K}$, where $\mathbf m$ comes out of the easy plane.  

Gyroscopic and potential forces acting on vortices are well understood. The gyroscopic density in (\ref{eq:G-Da}) comes from core regions, where $\mathbf m$ does not stay in a fixed plane \cite{Clarke2008}. Vortex cores are rigid objects, for which Thiele's approximation \cite{Thiele1973} works well. The gyroscopic force $\mathbf F^g$ for a vortex with velocity $\dot{\mathbf R} = (\dot{X}, \dot{Y})$ has components
\begin{equation}
F^g_{X} = - 4\pi Q \mathcal J h \dot{Y},
\quad
F^g_{Y} = 4\pi Q \mathcal J h \dot{X},
\label{eq:gyroscopic}
\end{equation}
where $Q = n p/2 = \pm 1/2$ is the skyrmion number of the vortex determined by its vorticity $n$ and polarity $p = \pm 1$. Exchange-mediated conservative forces between vortices resemble Coulomb interactions in two dimensions \cite{Kosterlitz1974}. The net conservative force on vortex $i$ is 
\begin{equation}
\mathbf F^c_i 
	= -2\pi A h \sum_{j \neq i} n_i n_j
    	\frac{\mathbf R_i - \mathbf R_j}
    		{|\mathbf R_i - \mathbf R_j|^2}.
\label{eq:Coulomb}
\end{equation}

Viscous forces are the primary focus of this paper. It is natural to expect that vortex $i$ experiences a viscous force $\mathbf F^v_i = - \hat{D}_{ii}\dot{\mathbf R}_i$, where $\hat{D}_{ii}$ is a $2\times2$ symmetric tensor with matrix elements such as $D_{X_i X_i}$, $D_{X_i Y_i}$, and so on. For magnetization lying primarily in the easy plane, $\mathbf m \approx (\cos{\phi}, \sin{\phi}, 0)$, Eq.~(\ref{eq:G-Db}) yields, e.g., 
\begin{equation}
D_{X_i X_i} 
	\approx \alpha |\mathcal J| h \int d^2r 
    	\left(
			\frac{\partial \phi}{\partial X_i}
        \right)^2
	= \alpha |\mathcal J| h \int d^2r 
    	\frac{(y-Y_i)^2}
        	{|\mathbf r - \mathbf R_i|^4}.
\nonumber
\end{equation}
On symmetry grounds, we expect $\hat{D}_{ii}$ to be isotropic, $\hat{D}_{ii} = D_{ii}\hat{1}$, with a scalar viscosity coefficient
\begin{equation}
D_{ii} = \frac{\alpha |\mathcal J| h}{2} \int 
    	\frac{d^2r}
        	{|\mathbf r - \mathbf R_i|^2}.
\label{eq:D-XiXi}
\end{equation}
The integral diverges and requires regularization for both $\mathbf r \to \mathbf R_i$ and $\mathbf r \to \infty$. The long-range cutoff is the system size \cite{Huber1982}; the short-range cutoff is provided by the size of the vortex core of the order $\lambda$ \cite{Sheka2005}. For a vortex near the center of a disk of radius $R_d \gg R_i$, 
\begin{equation}
D_{ii} \approx \alpha \pi |\mathcal J| h \ln{(R_d/C\lambda)},
\label{eq:D-ii}
\end{equation}
where $C$ is a numerical factor of the order 1. The logarithmic divergence with the system size $R_d$ reflects the long-range impact of a moving vortex on the magnetization distribution $\mathbf m(\mathbf r)$. Viscous forces in a ferromagnet are of the order $\alpha \ll 1$ (typically $10^{-4}$ to $10^{-2}$) and thus are much weaker than gyroscopic ones. For a vortex, this is partly compensated by the factor $\ln{(R_d/C\lambda)} \gg 1$. 

The extended nature of vortices leads to substantial \emph{mutual drag} between them. Vortex $i$ feels a force proportional to the velocity of vortex $j$, $\mathbf F^v_i = - \hat{D}_{ij} \dot{\mathbf R}_j$, where again $\hat{D}_{ij}$ is a $2\times2$ tensor with coefficients such as 
\begin{eqnarray*}
D_{X_i X_j} &\approx& 
	\alpha |\mathcal J| h \int d^2r \,
		\frac{\partial \phi}{\partial X_i}
    	\frac{\partial \phi}{\partial X_j}
\\
	&=& \alpha |\mathcal J| h n_i n_j
	\int d^2r \frac{(y-Y_i)(y-Y_j)}
    	{|\mathbf r - \mathbf R_i|^2
        	|\mathbf r - \mathbf R_j|^2}.
\end{eqnarray*}
The integrand has two singularities at $\mathbf r = \mathbf R_i$ and $\mathbf R_j$. They are weaker than the confluent singularity in Eq.~(\ref{eq:D-XiXi}) and are integrable, making the short-range cut-off $\lambda$ unnecessary. For two vortices located symmetrically about the center of a disk, $\mathbf R_i = (\pm R, 0)$ in Fig.~\ref{fig:v-av}, the viscosity tensor $\hat{D}_{ij}$ has principal axes parallel and perpendicular to the line connecting the cores with the eigenvalues 
\begin{subequations}
\begin{eqnarray}
D_{ij}^{||} &\approx& 
	\alpha \pi n_i n_j |\mathcal J| h
    	\left[
        	\ln{(R_d/2R)}
            + 1/2
        \right],
\label{eq:D-ija}
\\
D_{ij}^{\perp} &\approx& 
	\alpha \pi n_i n_j |\mathcal J| h
    	\left[
        	\ln{(R_d/2R)}
            - 1/2
        \right].
\label{eq:D-ijb}
\end{eqnarray}
\label{eq:D-ij}
\end{subequations}

\begin{table}[tbh]
\caption{Vortices in the numerical simulation.}
\label{tab:properties}
\begin{tabular}{||c|c|c|c|c|c|l||}
\hline\hline
$i$ & $n_i$ & $Q_i$ & $G_{X_i Y_i}$ & $R_i$ & $\Phi_i$ & note\\
\hline
1 & $+1$ & $+1/2$ & $-2\pi \mathcal J$ & $R$ & $\Phi$ 
	& vortex \\
2 & $-1$ & $+1/2$ & $-2\pi \mathcal J$ & $R$ & $\Phi + \pi$ 
	& antivortex \\
3 & $-1$ &  & 0 & $R_d^2/R$ & $\Phi$ 
	& image of the vortex \\
4 & $+1$ &  & 0 & $R_d^2/R$ & $\Phi + \pi$ 
	& image of the antivortex\\
\hline\hline
\end{tabular}
\end{table}

A notable feature of mutual drag is the dependence of its direction on vorticities. The drag force $\mathbf F^v_i = - \hat{D}_{ij} \dot{\mathbf R}_j$ on vortex $i$ is roughly opposite to the velocity $\dot{\mathbf R}_j$ of vortex $j$ for vorticities of the same sign and roughly parallel to it for vorticities of opposite sign. 

To test our theory, we modeled the dynamics of a vortex--antivortex pair with equal skyrmion numbers (Table~\ref{tab:properties}) with the aid of the micromagnetic simulator \textsc{MuMax3} \cite{Vansteenkiste2014}. We used magnetization length $\mathcal M = 8.60 \times 10^5 \ \mbox{A/m}$, gyromagnetic ratio $\gamma = -2.21 \times 10^5 \ \mbox{m/A s}$, angular momentum density $\mathcal J = \mu_0 \mathcal M/\gamma = - 4.89 \times 10^{-6}$ J s/m$^3$ \cite{sign-convention}, exchange constant $A = 2.6 \times 10^{-11} \ \mbox{J/m}$, and easy-plane anisotropy $K = 2.60 \times 10^5 \ \mbox{J/m}^3$. Natural units of length and time were $\lambda = \sqrt{A/K} = 10.0 \mbox{ nm}$ and $\tau = |\mathcal J|/K = 18.8$ ps. The sample was a disk of radius $R_D = 2048$ nm and thickness $h = 4$ nm. A vortex and an antivortex with equal skyrmion numbers $Q_1 = Q_2 = +1/2$ were initially placed symmetrically on opposite sides of the disk center, Fig.~\ref{fig:v-av}(a). The pair orbited the disk center and gradually spiraled down, Fig.~\ref{fig:v-av}(b). The dimensionless constant $C = 0.342$ in Eq.~(\ref{eq:D-ii}) was determined through a numerical evaluation of the dissipation constant $D_{ii} = \alpha |\mathcal J| h \int d^2r |\partial_x \mathbf m|^2$ of a simulated vortex. 

In the absence of dissipation ($\alpha=0$), the motion of the pair reflects the balance of the exchange-mediated attraction (\ref{eq:Coulomb}) and the gyroscopic force (\ref{eq:gyroscopic}), $\mathbf F^c + \mathbf F^g = 0$. The radial direction of the exchange attraction results in the azimuthal direction of the vortex velocities. The two topological defects orbit the common center at a constant radius $R$. It is therefore convenient to parametrize the positions of the vortices in polar coordinates $(R_i, \Phi_i)$, see Table~\ref{tab:properties}. The angular velocity is obtained from the balance of the gyroscopic and conservative forces acting on a vortex in the radial direction, 
\begin{equation}
-2\pi \mathcal J h R \dot{\Phi} - \frac{2\pi A h}{2 R} = 0.
\label{eq:eom-radial}
\end{equation}
Weak dissipation ($\alpha \ll 1$) turns the trajectories into spirals with a radial velocity $\dot{R}$ of the order $\alpha$. 

\begin{figure}[t!b]
\includegraphics[width=0.95\columnwidth]{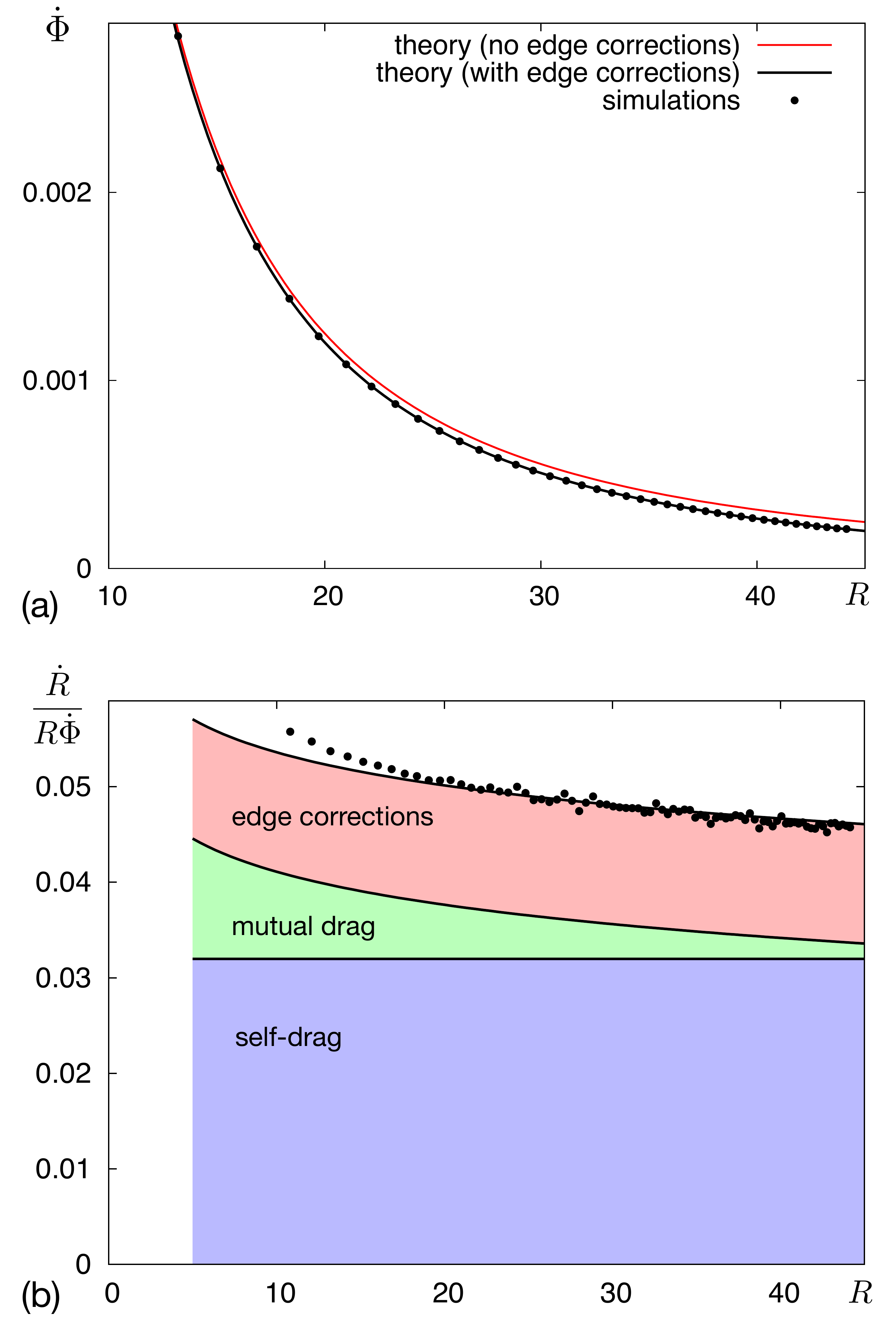}
\caption{(a) Angular velocity $\dot{\Phi}$ of the vortices vs. the radius $R$ of their orbit. (b) The ratio of radial $\dot{R}$ and orbital $R \dot{\Phi}$ velocities vs. the orbit radius $R$. Natural units of length $\lambda$ and time $\tau$.}
\label{fig:velocities}
\end{figure}

Numerical simulations reveal very good, but not perfect, agreement with Eq.~(\ref{eq:eom-radial}): the observed angular velocity of the vortices $\dot{\Phi}(R)$ differed from the expected value $\dot{\Phi} = -A/2 \mathcal J R^2$ by a small constant, Fig.~\ref{fig:velocities}(a). This minor discrepancy reflects an edge effect in a finite system. Free boundary conditions at the edge, $\partial \phi/\partial r = 0$, result in the appearance of image vortices outside of the disk \cite{Gaididei2000, Sheka2005}, see Table~\ref{tab:properties}. The images generate a weak radial force of approximately $4\pi A h R/R_d^2$ that reduces the angular velocity by $2A/\mathcal J R_d^2$, in excellent agreement with the numerical data, Fig.~\ref{fig:velocities}(a).

To determine the radial velocity $\dot{R}$, which is of the order $\alpha$, we need to carefully evaluate viscous forces acting on the vortices. This task is made complicated by the constrained motion of images (their positions mirror the locations of the vortices). Because of these constraints, the vortices also ``feel'' forces acting on the images. Although it is possible to solve the dynamics with constraints, a more expedient way is to reformulate the dynamics in terms of the two independent variables $R$ and $\Phi$ that fully determine the positions of all four objects (see Table~\ref{tab:properties}). Equations of motion for $R$ and $\Phi$ can be obtained by following the usual prescription (\ref{eq:CC}): 
\begin{subequations}
\begin{eqnarray}
G_{R\Phi} \dot{\Phi} - \partial U/\partial R - D_{RR} \dot{R} = 0,\label{eq:eom-R-Phia}\\
G_{\Phi R} \dot{R} - D_{\Phi\Phi} \dot{\Phi} = 0. \label{eq:eom-R-Phib}
\end{eqnarray}
\label{eq:eom-R-Phi}
\end{subequations}
Here we took into account rotational symmetry, which yields $-\partial U/\partial \Phi = 0$, and the diagonal nature of the dissipation tensor, $D_{R\Phi} = D_{\Phi R} = 0$. 

Polar components of the gyroscopic tensor $G_{R\Phi} = - G_{\Phi R}$ can be expressed in terms of Cartesian ones through a standard coordinate change from $\{q^\mu\}$ (here positions of the vortices and images $\mathbf R_i$) to $R$ and $\Phi$: 
\begin{eqnarray}
G_{R\Phi} &=& \sum_{\mu,\nu} 
	\frac{\partial q^\mu}{\partial R}
    \frac{\partial q^\nu}{\partial \Phi} 
    G_{q^\mu q^\nu} 
    = R(G_{X_1 Y_1} + G_{X_2 Y_2}) 
\nonumber\\
	&=& - 4\pi R \mathcal Jh .
\end{eqnarray}
Here we used the reference frame of Fig.~\ref{fig:v-av}, in which $\partial \mathbf R_1/\partial R = (1,0)$, $\partial \mathbf R_1/\partial \Phi = (0,R)$, $\partial \mathbf R_2/\partial R = (-1,0)$, $\partial \mathbf R_2/\partial \Phi = (0,-R)$, etc. Image vortices 3 and 4 do not contribute to the gyroscopic tensor because they are centered outside the sample and therefore lack cores. The energy of the vortices and images is 
\begin{equation}
U(R) = 2\pi A h 
	\left( 
    	\ln{2R} 
        + \ln{\frac{R_d^2 - R^2}{R_d^2 + R^2}}
    \right).
\end{equation}
The dissipative term in Eq.~(\ref{eq:eom-R-Phia}) is of the order $\alpha^2$ and can be neglected to yield $\dot{\Phi} \approx -(A/2\mathcal J)(R^{-2} - 4R_d^{-2})$ for $R \ll R_d$, as derived above. 

To obtain the radial velocity from Eq.~(\ref{eq:eom-R-Phib}), we need the dissipative coefficient 
\begin{equation}
D_{\Phi\Phi} = \sum_{\mu,\nu} 
	\frac{\partial q^\mu}{\partial \Phi}
    \frac{\partial q^\nu}{\partial \Phi} 
    D_{q^\mu q^\nu}
	= \sum_{i,j} (-1)^{i+j} R_i R_j D_{ij}^\perp,
\end{equation}
which reduces to a superposition of these terms:  
\begin{subequations}
\begin{eqnarray}
&&2R_1^2 D_{11} 
	\approx 
    2 \alpha \pi |\mathcal J| h R^2 
    \ln{(R_d/C \lambda)},
    \label{eq:D-alla}
\\
&&-2R_1^2 D_{12}^\perp
	\approx
    2 \alpha \pi |\mathcal J| h R^2
   		[\ln{(R_d/2 R)}-1/2],
 	\label{eq:D-allb}	
\\
&&4R_1 R_3(D_{13}^\perp - D_{14}^\perp)
	\approx 
    4 \alpha \pi |\mathcal J| h R^2,
   	\label{eq:D-allc}	
\\
&&2R_3^2(D_{33}^\perp - D_{34}^\perp)
	\approx 
    \alpha \pi |\mathcal J| h R^2.
    \label{eq:D-alld}
\end{eqnarray}
\label{eq:D-all}
\end{subequations}
They represent self-drag of the vortices (\ref{eq:D-alla}), their mutual drag (\ref{eq:D-allb}), and corrections from the images (\ref{eq:D-allc}) and (\ref{eq:D-alld}).  In the limit of a large disk, the first two terms dominate over the edge corrections, albeit only logarithmically in the system size $R_d$. Thus the edge corrections must be included to obtain quantitative agreement with simulations, Fig.~\ref{fig:velocities}(b). 

The ratio of the radial and orbital velocities, 
\begin{equation}
\frac{\dot{R}}{R \dot{\Phi}} 
	= \frac{D_{\Phi \Phi}}{R G_{\Phi R}}
	= - \mathrm{sgn}{\mathcal J} 
    	\frac{\alpha}{2} 
        \ln{\frac{\Lambda}{R}}, 
\end{equation}
where $\Lambda = R_d^2 e^2/2C \lambda$, has a telltale logarithmic dependence on the vortex separation $2R$ inherited from mutual drag, Eqs.~(\ref{eq:D-ijb}) and (\ref{eq:D-allb}).  As a result, a vortex-antivortex pair follows a \emph{double} logarithmic spiral, 
\begin{equation}
\ln{\ln{\frac{R(\Phi)}{\Lambda}}}
- \ln{\ln{\frac{R(0)}{\Lambda}}}
= - \mathrm{sgn}{\mathcal J} \frac{\alpha \Phi}{2}, 
\label{eq:double-log}
\end{equation}
in excellent agreement with the numerical simulations, Fig.~\ref{fig:v-av}(b). Note the contrast with a single vortex in a disk, which has a constant self-drag coefficient (\ref{eq:D-ii}) and therefore follows a simple logarithmic spiral \cite{Sheka2005}.

At the smallest orbital radii $R$, the observed radial velocity $\dot{R}$ shows small but growing deviations from the theoretical value, Fig.~\ref{fig:velocities}(b). As the radial motion is tied to energy dissipation, an excess radial velocity hints at the opening of a new dissipation channel. The likely culprit is spin waves, which have a linear dispersion $\omega = sk$ with the speed $s = \sqrt{AK/\mathcal J^2} = \lambda/\tau$. In a disk, the normal modes in polar coordinates $(R,\Phi)$ are
\begin{equation}
\phi(R,\Phi,t) = a J_m(kR) \cos{(\omega t - m \Phi)},
\end{equation}
where $J_m(x)$ is a Bessel function of the first kind. For open boundary conditions, the wavenumbers $k$ satisfy $J_m'(k R_d) = 0$. A rotating vortex-antivortex pair couples strongly to modes with $m = 1$. The lowest frequency for an $m = 1$ spin wave is $\omega = 1.84 s/R_d = 8.98 \times 10^{-3} \tau^{-1}$. At the end of the simulation, the angular frequency of the pair reached $\dot{\Phi} = 4.3 \times 10^{-3} \tau^{-1}$. Although the pair was not yet in resonance with this mode, its angular velocity had a substantial chirp, $|\ddot{\Phi}/\dot{\Phi}^2| = \alpha \ln{(\Lambda/R)} \approx 0.1$, and thus a spectrum potentially wide enough to excite the $m=1$ spin wave, whose dynamics would produce additional dissipation. 

We have derived viscous forces acting on a vortex in a thin-film ferromagnet. In addition to self-drag proportional to the vortex's own velocity, vortices experience mutual drag, a force on vortex $i$ proportional to the velocity of vortex $j$, $\mathbf F^v_i = - \hat{D}_{ij} \dot{\mathbf R}_j$. Reflecting the long-range influence of vortices, both the self-drag coefficient $D_{ii}$ (\ref{eq:D-ii}) and the mutual viscosity tensor $\hat{D}_{ij}$ (\ref{eq:D-ij}) scale logarithmically with the system size. The mutual drag tensor $\hat{D}_{ij}$ is anisotropic and distance-dependent. We have tested our theory by deriving the dynamics of a vortex-antivortex pair with equal skyrmion numbers. The results are in excellent agreement with micromagnetic simulations. Edge effects in the form of image vortices contribute substantially to viscous friction. Dissipation through the emission of spin waves becomes noticeable when vortices approach each other very closely, within a few exchange lengths $\lambda$. 

The authors thank Se Kwon Kim for helpful discussions. This work was supported by the US Department of Energy, Office of Basic Energy Sciences, Division of Materials Sciences and Engineering under Award DE-FG02-08ER46544.


\bibliographystyle{apsrev4-1}
\bibliography{main}

\begin{thebibliography}{21}%
\makeatletter
\providecommand \@ifxundefined [1]{%
 \@ifx{#1\undefined}
}%
\providecommand \@ifnum [1]{%
 \ifnum #1\expandafter \@firstoftwo
 \else \expandafter \@secondoftwo
 \fi
}%
\providecommand \@ifx [1]{%
 \ifx #1\expandafter \@firstoftwo
 \else \expandafter \@secondoftwo
 \fi
}%
\providecommand \natexlab [1]{#1}%
\providecommand \enquote  [1]{``#1''}%
\providecommand \bibnamefont  [1]{#1}%
\providecommand \bibfnamefont [1]{#1}%
\providecommand \citenamefont [1]{#1}%
\providecommand \href@noop [0]{\@secondoftwo}%
\providecommand \href [0]{\begingroup \@sanitize@url \@href}%
\providecommand \@href[1]{\@@startlink{#1}\@@href}%
\providecommand \@@href[1]{\endgroup#1\@@endlink}%
\providecommand \@sanitize@url [0]{\catcode `\\12\catcode `\$12\catcode
  `\&12\catcode `\#12\catcode `\^12\catcode `\_12\catcode `\%12\relax}%
\providecommand \@@startlink[1]{}%
\providecommand \@@endlink[0]{}%
\providecommand \url  [0]{\begingroup\@sanitize@url \@url }%
\providecommand \@url [1]{\endgroup\@href {#1}{\urlprefix }}%
\providecommand \urlprefix  [0]{URL }%
\providecommand \Eprint [0]{\href }%
\providecommand \doibase [0]{http://dx.doi.org/}%
\providecommand \selectlanguage [0]{\@gobble}%
\providecommand \bibinfo  [0]{\@secondoftwo}%
\providecommand \bibfield  [0]{\@secondoftwo}%
\providecommand \translation [1]{[#1]}%
\providecommand \BibitemOpen [0]{}%
\providecommand \bibitemStop [0]{}%
\providecommand \bibitemNoStop [0]{.\EOS\space}%
\providecommand \EOS [0]{\spacefactor3000\relax}%
\providecommand \BibitemShut  [1]{\csname bibitem#1\endcsname}%
\let\auto@bib@innerbib\@empty
\bibitem [{sig()}]{sign-convention}%
  \BibitemOpen
  \href@noop {} {}\bibinfo {note} {Here $\mathcal J = \mu_0\mathcal M/\gamma$,
  where $\mathcal M > 0$ is magnetization length and $\gamma$ is the
  gyromagnetic ratio. Thus $\mathrm{sgn}\mathcal J = \mathrm{sgn}
  \gamma$.}\BibitemShut {Stop}%
\bibitem [{\citenamefont {Gilbert}(2004)}]{Gilbert2004}%
  \BibitemOpen
  \bibfield  {author} {\bibinfo {author} {\bibfnamefont {T.~L.}\ \bibnamefont
  {Gilbert}},\ }\href {\doibase 10.1109/TMAG.2004.836740} {\bibfield  {journal}
  {\bibinfo  {journal} {IEEE Trans. Mag.}\ }\textbf {\bibinfo {volume} {40}},\
  \bibinfo {pages} {3443} (\bibinfo {year} {2004})}\BibitemShut {NoStop}%
\bibitem [{\citenamefont {Thiele}(1973)}]{Thiele1973}%
  \BibitemOpen
  \bibfield  {author} {\bibinfo {author} {\bibfnamefont {A.~A.}\ \bibnamefont
  {Thiele}},\ }\href {\doibase 10.1103/PhysRevLett.30.230} {\bibfield
  {journal} {\bibinfo  {journal} {Phys. Rev. Lett.}\ }\textbf {\bibinfo
  {volume} {30}},\ \bibinfo {pages} {230} (\bibinfo {year} {1973})}\BibitemShut
  {NoStop}%
\bibitem [{\citenamefont {Takeno}\ and\ \citenamefont
  {Homma}(1982)}]{Takeno1982}%
  \BibitemOpen
  \bibfield  {author} {\bibinfo {author} {\bibfnamefont {S.}~\bibnamefont
  {Takeno}}\ and\ \bibinfo {author} {\bibfnamefont {S.}~\bibnamefont {Homma}},\
  }\href {\doibase 10.1143/PTP.67.1633} {\bibfield  {journal} {\bibinfo
  {journal} {Prog. Theor. Phys.}\ }\textbf {\bibinfo {volume} {67}},\ \bibinfo
  {pages} {1633} (\bibinfo {year} {1982})}\BibitemShut {NoStop}%
\bibitem [{\citenamefont {Huber}(1982)}]{Huber1982}%
  \BibitemOpen
  \bibfield  {author} {\bibinfo {author} {\bibfnamefont {D.~L.}\ \bibnamefont
  {Huber}},\ }\href {\doibase 10.1103/PhysRevB.26.3758} {\bibfield  {journal}
  {\bibinfo  {journal} {Phys. Rev. B}\ }\textbf {\bibinfo {volume} {26}},\
  \bibinfo {pages} {3758} (\bibinfo {year} {1982})}\BibitemShut {NoStop}%
\bibitem [{\citenamefont {Nikiforov}\ and\ \citenamefont
  {Sonin}(1983)}]{Nikiforov1983}%
  \BibitemOpen
  \bibfield  {author} {\bibinfo {author} {\bibfnamefont {A.~V.}\ \bibnamefont
  {Nikiforov}}\ and\ \bibinfo {author} {\bibfnamefont {{\'E}.~B.}\ \bibnamefont
  {Sonin}},\ }\href@noop {} {\bibfield  {journal} {\bibinfo  {journal} {Sov.
  Phys. JETP}\ }\textbf {\bibinfo {volume} {58}},\ \bibinfo {pages} {373}
  (\bibinfo {year} {1983})}\BibitemShut {NoStop}%
\bibitem [{\citenamefont {Mertens}\ \emph {et~al.}(1997)\citenamefont
  {Mertens}, \citenamefont {Schnitzer},\ and\ \citenamefont
  {Bishop}}]{Mertens1997}%
  \BibitemOpen
  \bibfield  {author} {\bibinfo {author} {\bibfnamefont {F.~G.}\ \bibnamefont
  {Mertens}}, \bibinfo {author} {\bibfnamefont {H.~J.}\ \bibnamefont
  {Schnitzer}}, \ and\ \bibinfo {author} {\bibfnamefont {A.~R.}\ \bibnamefont
  {Bishop}},\ }\href {\doibase 10.1103/PhysRevB.56.2510} {\bibfield  {journal}
  {\bibinfo  {journal} {Phys. Rev. B}\ }\textbf {\bibinfo {volume} {56}},\
  \bibinfo {pages} {2510} (\bibinfo {year} {1997})}\BibitemShut {NoStop}%
\bibitem [{\citenamefont {Mertens}\ and\ \citenamefont
  {Bishop}(2000)}]{Mertens2000}%
  \BibitemOpen
  \bibfield  {author} {\bibinfo {author} {\bibfnamefont {F.~G.}\ \bibnamefont
  {Mertens}}\ and\ \bibinfo {author} {\bibfnamefont {A.~R.}\ \bibnamefont
  {Bishop}},\ }in\ \href {\doibase 10.1007/3-540-46629-0} {\emph {\bibinfo
  {booktitle} {Nonlinear Science at the Dawn of the 21th Century}}},\ \bibinfo
  {series and number} {Lecture Notes in Physics},\ \bibinfo {editor} {edited
  by\ \bibinfo {editor} {\bibfnamefont {P.~L.}\ \bibnamefont {Christiansen}},
  \bibinfo {editor} {\bibfnamefont {M.~P.}\ \bibnamefont {Soerensen}}, \ and\
  \bibinfo {editor} {\bibfnamefont {A.~C.}\ \bibnamefont {Scott}}}\ (\bibinfo
  {publisher} {Springer},\ \bibinfo {address} {Berlin},\ \bibinfo {year}
  {2000})\ pp.\ \bibinfo {pages} {137--170}\BibitemShut {NoStop}%
\bibitem [{\citenamefont {Gaididei}\ \emph {et~al.}(2000)\citenamefont
  {Gaididei}, \citenamefont {Kamppeter}, \citenamefont {Mertens},\ and\
  \citenamefont {Bishop}}]{Gaididei2000}%
  \BibitemOpen
  \bibfield  {author} {\bibinfo {author} {\bibfnamefont {Y.}~\bibnamefont
  {Gaididei}}, \bibinfo {author} {\bibfnamefont {T.}~\bibnamefont {Kamppeter}},
  \bibinfo {author} {\bibfnamefont {F.~G.}\ \bibnamefont {Mertens}}, \ and\
  \bibinfo {author} {\bibfnamefont {A.~R.}\ \bibnamefont {Bishop}},\ }\href
  {\doibase 10.1103/PhysRevB.61.9449} {\bibfield  {journal} {\bibinfo
  {journal} {Phys. Rev. B}\ }\textbf {\bibinfo {volume} {61}},\ \bibinfo
  {pages} {9449} (\bibinfo {year} {2000})}\BibitemShut {NoStop}%
\bibitem [{\citenamefont {Guslienko}\ \emph {et~al.}(2002)\citenamefont
  {Guslienko}, \citenamefont {Ivanov}, \citenamefont {Novosad}, \citenamefont
  {Otani}, \citenamefont {Shima},\ and\ \citenamefont
  {Fukamichi}}]{Guslienko2002}%
  \BibitemOpen
  \bibfield  {author} {\bibinfo {author} {\bibfnamefont {K.~Y.}\ \bibnamefont
  {Guslienko}}, \bibinfo {author} {\bibfnamefont {B.~A.}\ \bibnamefont
  {Ivanov}}, \bibinfo {author} {\bibfnamefont {V.}~\bibnamefont {Novosad}},
  \bibinfo {author} {\bibfnamefont {Y.}~\bibnamefont {Otani}}, \bibinfo
  {author} {\bibfnamefont {H.}~\bibnamefont {Shima}}, \ and\ \bibinfo {author}
  {\bibfnamefont {K.}~\bibnamefont {Fukamichi}},\ }\href {\doibase
  10.1063/1.1450816} {\bibfield  {journal} {\bibinfo  {journal} {J. Appl.
  Phys.}\ }\textbf {\bibinfo {volume} {91}},\ \bibinfo {pages} {8037} (\bibinfo
  {year} {2002})}\BibitemShut {NoStop}%
\bibitem [{\citenamefont {Kovalev}\ \emph {et~al.}(2002)\citenamefont
  {Kovalev}, \citenamefont {Komineas},\ and\ \citenamefont
  {Mertens}}]{Kovalev2002}%
  \BibitemOpen
  \bibfield  {author} {\bibinfo {author} {\bibfnamefont {A.}~\bibnamefont
  {Kovalev}}, \bibinfo {author} {\bibfnamefont {S.}~\bibnamefont {Komineas}}, \
  and\ \bibinfo {author} {\bibfnamefont {F.}~\bibnamefont {Mertens}},\ }\href
  {\doibase 10.1140/e10051-002-0010-1} {\bibfield  {journal} {\bibinfo
  {journal} {Eur. J. Phys. B}\ }\textbf {\bibinfo {volume} {25}},\ \bibinfo
  {pages} {89} (\bibinfo {year} {2002})}\BibitemShut {NoStop}%
\bibitem [{\citenamefont {Buchanan}\ \emph {et~al.}(2005)\citenamefont
  {Buchanan}, \citenamefont {Roy}, \citenamefont {Grimsditch}, \citenamefont
  {Fradin}, \citenamefont {Guslienko}, \citenamefont {Bader},\ and\
  \citenamefont {Novosad}}]{Buchanan2005}%
  \BibitemOpen
  \bibfield  {author} {\bibinfo {author} {\bibfnamefont {K.~S.}\ \bibnamefont
  {Buchanan}}, \bibinfo {author} {\bibfnamefont {P.~E.}\ \bibnamefont {Roy}},
  \bibinfo {author} {\bibfnamefont {M.}~\bibnamefont {Grimsditch}}, \bibinfo
  {author} {\bibfnamefont {F.~Y.}\ \bibnamefont {Fradin}}, \bibinfo {author}
  {\bibfnamefont {K.~Y.}\ \bibnamefont {Guslienko}}, \bibinfo {author}
  {\bibfnamefont {S.~D.}\ \bibnamefont {Bader}}, \ and\ \bibinfo {author}
  {\bibfnamefont {V.}~\bibnamefont {Novosad}},\ }\href {\doibase
  10.1038/nphys173} {\bibfield  {journal} {\bibinfo  {journal} {Nat. Phys.}\
  }\textbf {\bibinfo {volume} {1}},\ \bibinfo {pages} {172} (\bibinfo {year}
  {2005})}\BibitemShut {NoStop}%
\bibitem [{\citenamefont {Sheka}\ \emph {et~al.}(2005)\citenamefont {Sheka},
  \citenamefont {Zagorodny}, \citenamefont {Caputo}, \citenamefont {Gaididei},\
  and\ \citenamefont {Mertens}}]{Sheka2005}%
  \BibitemOpen
  \bibfield  {author} {\bibinfo {author} {\bibfnamefont {D.~D.}\ \bibnamefont
  {Sheka}}, \bibinfo {author} {\bibfnamefont {J.~P.}\ \bibnamefont
  {Zagorodny}}, \bibinfo {author} {\bibfnamefont {J.-G.}\ \bibnamefont
  {Caputo}}, \bibinfo {author} {\bibfnamefont {Y.}~\bibnamefont {Gaididei}}, \
  and\ \bibinfo {author} {\bibfnamefont {F.~G.}\ \bibnamefont {Mertens}},\
  }\href {\doibase 10.1103/PhysRevB.71.134420} {\bibfield  {journal} {\bibinfo
  {journal} {Phys. Rev. B}\ }\textbf {\bibinfo {volume} {71}},\ \bibinfo
  {pages} {134420} (\bibinfo {year} {2005})}\BibitemShut {NoStop}%
\bibitem [{\citenamefont {Komineas}(2007)}]{Komineas2007}%
  \BibitemOpen
  \bibfield  {author} {\bibinfo {author} {\bibfnamefont {S.}~\bibnamefont
  {Komineas}},\ }\href {\doibase 10.1103/PhysRevLett.99.117202} {\bibfield
  {journal} {\bibinfo  {journal} {Phys. Rev. Lett.}\ }\textbf {\bibinfo
  {volume} {99}},\ \bibinfo {pages} {117202} (\bibinfo {year}
  {2007})}\BibitemShut {NoStop}%
\bibitem [{\citenamefont {Ivanov}\ \emph {et~al.}(2010)\citenamefont {Ivanov},
  \citenamefont {Galkina},\ and\ \citenamefont {Galkin}}]{Ivanov2010}%
  \BibitemOpen
  \bibfield  {author} {\bibinfo {author} {\bibfnamefont {B.~A.}\ \bibnamefont
  {Ivanov}}, \bibinfo {author} {\bibfnamefont {E.~G.}\ \bibnamefont {Galkina}},
  \ and\ \bibinfo {author} {\bibfnamefont {A.~Y.}\ \bibnamefont {Galkin}},\
  }\href {\doibase 10.1063/1.3490861} {\bibfield  {journal} {\bibinfo
  {journal} {Low Temp. Phys.}\ }\textbf {\bibinfo {volume} {36}},\ \bibinfo
  {pages} {747} (\bibinfo {year} {2010})}\BibitemShut {NoStop}%
\bibitem [{\citenamefont {Tretiakov}\ \emph {et~al.}(2008)\citenamefont
  {Tretiakov}, \citenamefont {Clarke}, \citenamefont {Chern}, \citenamefont
  {Bazaliy},\ and\ \citenamefont {Tchernyshyov}}]{Tretiakov2008}%
  \BibitemOpen
  \bibfield  {author} {\bibinfo {author} {\bibfnamefont {O.~A.}\ \bibnamefont
  {Tretiakov}}, \bibinfo {author} {\bibfnamefont {D.}~\bibnamefont {Clarke}},
  \bibinfo {author} {\bibfnamefont {G.-W.}\ \bibnamefont {Chern}}, \bibinfo
  {author} {\bibfnamefont {Y.~B.}\ \bibnamefont {Bazaliy}}, \ and\ \bibinfo
  {author} {\bibfnamefont {O.}~\bibnamefont {Tchernyshyov}},\ }\href {\doibase
  10.1103/PhysRevLett.100.127204} {\bibfield  {journal} {\bibinfo  {journal}
  {Phys. Rev. Lett.}\ }\textbf {\bibinfo {volume} {100}},\ \bibinfo {pages}
  {127204} (\bibinfo {year} {2008})}\BibitemShut {NoStop}%
\bibitem [{\citenamefont {Clarke}\ \emph {et~al.}(2008)\citenamefont {Clarke},
  \citenamefont {Tretiakov}, \citenamefont {Chern}, \citenamefont {Bazaliy},\
  and\ \citenamefont {Tchernyshyov}}]{Clarke2008}%
  \BibitemOpen
  \bibfield  {author} {\bibinfo {author} {\bibfnamefont {D.~J.}\ \bibnamefont
  {Clarke}}, \bibinfo {author} {\bibfnamefont {O.~A.}\ \bibnamefont
  {Tretiakov}}, \bibinfo {author} {\bibfnamefont {G.-W.}\ \bibnamefont
  {Chern}}, \bibinfo {author} {\bibfnamefont {Y.~B.}\ \bibnamefont {Bazaliy}},
  \ and\ \bibinfo {author} {\bibfnamefont {O.}~\bibnamefont {Tchernyshyov}},\
  }\href {\doibase 10.1103/PhysRevB.78.134412} {\bibfield  {journal} {\bibinfo
  {journal} {Phys. Rev. B}\ }\textbf {\bibinfo {volume} {78}},\ \bibinfo
  {pages} {134412} (\bibinfo {year} {2008})}\BibitemShut {NoStop}%
\bibitem [{dip()}]{dipolar}%
  \BibitemOpen
  \href@noop {} {}\bibinfo {note} {To be exact, effects of dipolar interactions
  are partly included in our model. In a thin film, the anisotropy constant $K$
  includes a dipolar contribution $\mu_0\mathcal M^2/2$. We neglect the
  nonlocal stray-field effects.}\BibitemShut {Stop}%
\bibitem [{\citenamefont {Chaikin}\ and\ \citenamefont
  {Lubensky}(2000)}]{Chaikin-Lubensky}%
  \BibitemOpen
  \bibfield  {author} {\bibinfo {author} {\bibfnamefont {P.~M.}\ \bibnamefont
  {Chaikin}}\ and\ \bibinfo {author} {\bibfnamefont {T.~C.}\ \bibnamefont
  {Lubensky}},\ }\href@noop {} {\emph {\bibinfo {title} {Principles of
  Condensed Matter Physics}}}\ (\bibinfo  {publisher} {Cambridge University
  Press},\ \bibinfo {year} {2000})\BibitemShut {NoStop}%
\bibitem [{\citenamefont {Kosterlitz}(1974)}]{Kosterlitz1974}%
  \BibitemOpen
  \bibfield  {author} {\bibinfo {author} {\bibfnamefont {J.~M.}\ \bibnamefont
  {Kosterlitz}},\ }\href {\doibase 10.1088/0022-3719/7/6/005} {\bibfield
  {journal} {\bibinfo  {journal} {J. Phys. C}\ }\textbf {\bibinfo {volume}
  {7}},\ \bibinfo {pages} {1046} (\bibinfo {year} {1974})}\BibitemShut
  {NoStop}%
\bibitem [{\citenamefont {Vansteenkiste}\ \emph {et~al.}(2014)\citenamefont
  {Vansteenkiste}, \citenamefont {Leliaert}, \citenamefont {Dvornik},
  \citenamefont {Helsen}, \citenamefont {Garcia-Sanchez},\ and\ \citenamefont
  {Van~Waeyenberge}}]{Vansteenkiste2014}%
  \BibitemOpen
  \bibfield  {author} {\bibinfo {author} {\bibfnamefont {A.}~\bibnamefont
  {Vansteenkiste}}, \bibinfo {author} {\bibfnamefont {J.}~\bibnamefont
  {Leliaert}}, \bibinfo {author} {\bibfnamefont {M.}~\bibnamefont {Dvornik}},
  \bibinfo {author} {\bibfnamefont {M.}~\bibnamefont {Helsen}}, \bibinfo
  {author} {\bibfnamefont {F.}~\bibnamefont {Garcia-Sanchez}}, \ and\ \bibinfo
  {author} {\bibfnamefont {B.}~\bibnamefont {Van~Waeyenberge}},\ }\href
  {\doibase 10.1063/1.4899186} {\bibfield  {journal} {\bibinfo  {journal} {AIP
  Adv.}\ }\textbf {\bibinfo {volume} {4}},\ \bibinfo {pages} {107133} (\bibinfo
  {year} {2014})}\BibitemShut {NoStop}%
\end{thebibliography}%

\end{document}